\documentclass[aps,prb,twocolumn,superscriptaddress]{revtex4}
\usepackage{graphicx}
\usepackage{array}
\usepackage{amssymb}
\usepackage{txfonts}
\usepackage{textcomp}
\usepackage{color}

\begin{document}

%\preprint{}
\title{Nematic fluctuations in the non-superconducting iron pnictide BaFe$_{1.9-x}$Ni$_{0.1}$Cr$_{x}$As$_{2}$}

\author{Dongliang Gong}
\thanks{Present address: Department of Physics and Astronomy, University of Tennessee, Knoxville TN 37996, USA}
\affiliation{Beijing National Laboratory for Condensed Matter Physics, Institute of Physics, Chinese Academy of Sciences, Beijing
100190, China}
\affiliation{University of Chinese Academy of Sciences, Beijing 100049, China}

\author{Ming Yi}
\affiliation{Department of Physics and Astronomy, Rice University, Houston, Texas 77005, USA}
\affiliation{Department of Physics, University of California, Berkeley, California 94720, USA}

\author{Meng Wang}
\affiliation{School of Physics, Sun Yat-Sen University, Guangzhou 510275, China}

\author{Tao Xie}
\thanks{Present address: Neutron Scattering Division, Oak Ridge National Laboratory, Oak Ridge, Tennessee 37831, USA}
\affiliation{Beijing National Laboratory for Condensed Matter Physics, Institute of Physics, Chinese Academy of Sciences, Beijing
100190, China}
\affiliation{University of Chinese Academy of Sciences, Beijing 100049, China}

\author{Wenliang Zhang}
\thanks{Present address: Photon Science Division, Swiss Light Source, Paul Scherrer Institut, CH-5232 Villigen PSI, Switzerland}
\affiliation{Beijing National Laboratory for Condensed Matter Physics, Institute of Physics, Chinese Academy of Sciences, Beijing
100190, China}
\affiliation{University of Chinese Academy of Sciences, Beijing 100049, China}

\author{Sergey Danilkin}
\affiliation{Australian Centre for Neutron Scattering, Australian Nuclear Science and
Technology Organisation, Lucas Heights NSW-2234, Australia}

\author{Guochu Deng}
\affiliation{Australian Centre for Neutron Scattering, Australian Nuclear Science and
Technology Organisation, Lucas Heights NSW-2234, Australia}

\author{Xinzhi Liu}
\affiliation{School of Physics, Sun Yat-Sen University, Guangzhou 510275, China}

\author{Jitae T. Park}
\affiliation{Heinz Maier-Leibnitz Zentrum (MLZ), Technische Universit\"{a}t M\"{u}nchen, D-85748 Garching, Germany}

\author{Kazuhiko Ikeuchi}
\affiliation{Neutron Science and Technology Center, Comprehensive Research Organization for Science and Society,
Tokai, Ibaraki 319-1106, Japan}

\author{Kazuya Kamazawa}
\affiliation{Neutron Science and Technology Center, Comprehensive Research Organization for Science and Society,
Tokai, Ibaraki 319-1106, Japan}

\author{Sung-Kwan Mo}
\affiliation{Lawrence Berkeley National Laboratory, Berkeley, 94720 California, USA}

\author{Makoto Hashimoto}
\affiliation{Stanford Synchrotron Radiation Lightsource, SLAC National Accelerator Laboratory, Menlo
Park, California 94025, USA}

\author{Donghui Lu}
\affiliation{Stanford Synchrotron Radiation Lightsource, SLAC National Accelerator Laboratory, Menlo
Park, California 94025, USA}

\author{Rui Zhang}
\affiliation{Department of Physics and Astronomy, Rice Center for Quantum Materials, Rice University,
Houston, Texas 77005, USA}

\author{Pengcheng Dai}
\affiliation{Department of Physics and Astronomy, Rice Center for Quantum Materials, Rice University,
Houston, Texas 77005, USA}

\author{Robert J. Birgeneau}
\affiliation{Department of Physics, University of California, Berkeley, California 94720, USA}
\affiliation{Materials Science Division, Lawrence Berkeley National Laboratory, Berkeley, California 94720, USA}
\affiliation{Department of Materials Science and Engineering, University of California, Berkeley, California 94720, USA}

\author{Shiliang Li}
\affiliation{Beijing National Laboratory for Condensed Matter Physics, Institute of Physics, Chinese Academy of Sciences, Beijing
100190, China}
\affiliation{University of Chinese Academy of Sciences, Beijing 100049, China}
\affiliation{Songshan Lake Materials Laboratory, Dongguan, Guangdong 523808, China}

\author{Huiqian Luo}
\email{hqluo@iphy.ac.cn}
\affiliation{Beijing National Laboratory for Condensed Matter Physics, Institute of Physics, Chinese Academy of Sciences, Beijing
100190, China}
\affiliation{Songshan Lake Materials Laboratory, Dongguan, Guangdong 523808, China}

\begin{abstract}
The main driven force of the electronic nematic phase in iron-based superconductors is still under debate. Here, we report a comprehensive study on the nematic fluctuations in a non-superconducting iron pnictide system BaFe$_{1.9-x}$Ni$_{0.1}$Cr$_{x}$As$_{2}$ by electronic transport, angle-resolved photoemission spectroscopy (ARPES) and inelastic neutron scattering (INS) measurements. Previous neutron diffraction and transport measurements suggested that the collinear antiferromagnetism persists to $x=0.8$, with similar N\'{e}el temperature $T_N$ and structural transition temperature $T_s$ around 32 K, but the charge carriers change from electron type to hole type around $x=$ 0.5. In this study, we have found that the in-plane resistivity anisotropy also highly depends on the Cr dopings and the type of charge carriers. While ARPES measurements suggest possibly weak orbital anisotropy onset near $T_s$ for both $x=0.05$ and $x=0.5$ compounds, INS experiments reveal clearly different onset temperatures of low-energy spin excitation anisotropy, which is likely related to the energy scale of spin nematicity. These results suggest that the interplay between the local spins on Fe atoms and the itinerant electrons on Fermi surfaces is crucial to the nematic fluctuations of iron pnictides, where the orbital degree of freedom may behave differently from the spin degree of freedom, and the transport properties are intimately related to the spin dynamics.
\end{abstract}


\maketitle

\section{Introduction}

Electronic nematic phase breaks the rotational symmetry but preserves the translational symmetry of the underlying lattice in correlated materials \cite{oganesyan2001,fradkin2010a,fradkin2010b,wwang2021}. In iron-based superconductors, the nematic order associated with a tetragonal-to-orthorhombic structural transition at temperature $T_s$ acts as a precursor of the magnetic order below $T_N$ and the superconducting state below $T_c$ \cite{fernandes2012,fernandes2017,chen2014,pdai2015,si2016,gong2018}. The nematic fluctuations can be described by the electronic nematic susceptibility, which is defined as the susceptibility of electronic anisotropy to the uniaxial in-plane strain \cite{bohmer2016}. Divergent nematic susceptibility upon approaching $T_s$ from high temperature are revealed by the elastoresistance and elastic moduli measurements, suggesting nematic fluctuations well above $T_s$ \cite{jhchu2012,hhkuo2014,hhkuo2013,bohmer2014,dgong2017}.  The nematic fluctuations commonly exist in iron-based superconductors, and are even present in compounds with tetragonal crystal symmetry without any static nematic order \cite{bohmer2020}. Accumulating evidence suggests that the optimal superconductivity with maximum $T_c$ usually occurs near a nematic quantum critical point where the nematic fluctuations are the strongest\cite{hhkuo2016,yshizawa2012,jdai2009,skasahara2012,tshibauchi2013,slderer2015,hluo2012,xlu2013,dhu2015,wzhang2019,zliu2016,ygu2017}. However, the charge, spin and orbital degrees of freedom are always intertwined in the presence of nematic fluctuations \cite{chandra1990,jphu2012,fernandes2014,fernandes2012b,fwang2015,cma2014,cthorsmolle2014,qswang2015,chubukov2015,yamakawa2016}, giving a twofold rotational ($C_2$) symmetry in many physical properties \cite{fernandes2012,fernandes2017,chen2014, pdai2015,si2016,gong2018,bohmer2016,cclee2009,kruger2009,wclv2009,cchen2010,valenzuela2010} including anisotropic in-plane electronic resistivity and optical conductivity \cite{jhchu2010,matanatar2010,jjying2011,hman2015,xluo2015,cmirri2014,fisher2011}, lifting of degeneracy between $d_{xz}$/$d_{yz}$ orbitals \cite{myi2011,myi2014,myi2012,myi2019,yzhang2012,myi2017,watson2019}, anisotropic spin excitations at low energies \cite{hluo2013,xlu2014,wzhang2016,ysong2015,xlu2018}, phonon-energy split in lattice dynamics \cite{xren2015,ywhu2016}, and splitting of the Knight shift \cite{baek2014,tiye2015}.  In addition, it has been proposed that the local anisotropic impurity scattering of chemical dopants likely induces the twofold symmetry in the transport properties \cite{rosenthal2014,sishida2013,allan2013}. Such complex cases make it is difficult to clarify the main driven force of nematic phase by a single experimental probe.

Our previous works suggest that the Cr substitution is an effective way both to suppress the superconductivity and to tune the magnetism in iron-based superconductors \cite{wzhang2019,rzhang2014,rzhang2015,dgong2018}. Specifically in the BaFe$_{1.9-x}$Ni$_{0.1}$Cr$_{x}$As$_{2}$ system, by continuously doping Cr to the optimally superconducting compound BaFe$_{1.9}$Ni$_{0.1}$As$_2$ with $T_c=20$ K, the superconductivity is quickly suppressed above $x=0.05$, but the magnetic transition temperature $T_N$ and the structural transition temperature $T_s$ remain between 30 K and 35 K as shown by neutron diffraction results on naturally twinned samples [Fig. 1(a)]. Moreover, the effective moment $m$ is significantly enhanced first then suppressed for dopings higher than $x=0.5$, where the charge carriers change from electron type to hole type as shown by the sign of Hall and Seebeck coefficients \cite{dgong2018}. These make BaFe$_{1.9-x}$Ni$_{0.1}$Cr$_{x}$As$_{2}$ a rare example to separately tune the magnetically ordered temperature $T_N$ by the local spin interactions and the magnetically ordered strength by the scattering of itinerant electrons on Fermi surfaces, respectively. The extra holes introduced by Cr substitutions compensate the electron doping thus may drive those non-superconducting compounds to a half-filled Mott insulator similar to the parent compounds of cuprate and nickelate superconductors \cite{jmpizarro2017,medelmann2017,medici2014,paLee2014,qgu2022,ysong2016b}.  It would be interesting to monitor the evolution of the nematic fluctuations starting from a metallic state toward to a localized insulating state \cite{zpyin2011,georges2013,ysong2016b}, especially on the detwinned samples [Fig. 1(b)].

In this paper, we further report a multi-probe study on the nematic fluctuations in the non-superconducting compounds BaFe$_{1.9-x}$Ni$_{0.1}$Cr$_{x}$As$_{2}$ ($x=0.05 \sim 0.8$) by electronic transport, angle resolved photoemission spectroscopy (ARPES) and inelastic neutron scattering (INS) measurements. The in-plane resistivity anisotropy measured in the detwinned samples under uniaxial pressure shows a strong dependence on the Cr content with a clear sign change above $x=0.6$. By focusing on two compounds with $x=0.05$ and 0.5, ARPES measurements suggest possible band shifts induced by orbital anisotropy near $T_s$/$T_N$ for both dopings, but INS experiments reveal clearly different behaviors on the spin nematicity. The onset temperature of low-energy spin excitation anisotropy between $Q=(1, 0, 1)$ and $Q=(0, 1, 1)$ for $x=0.05$ is about 110 K, but for $x=0.5$ is much lower, only about 35 K near the magnetic transition. Such temperature dependence of spin nematicity is consistent with the results of in-plane resistivity anisotropy. At high energies, the spin nematicity for $x=0.05$ extends to about 120 meV, much larger than the case for $x=0.5$ (about 40 meV), suggesting a possible linear correlation between the highest energy scale and the onset temperature of spin nematicity. Therefore, the nematic behaviors in iron pnictides is highly related to the interplay between local moments and itinerant electrons. While the $C_2$-type anisotropies in spin excitations and in-plane resistivity are strongly correlated to each other \cite{xlu2014}, the orbital anisotropy induced band splitting may behave differently as affected by the complex Fermi surface topology \cite{myi2009,richard2011,ysong2017,ysong2016,txie2018a,txie2018b,txie2020,twang2020a,jguo2019,cliu2022}.

\begin{figure}[h]
\includegraphics[width=0.47\textwidth]{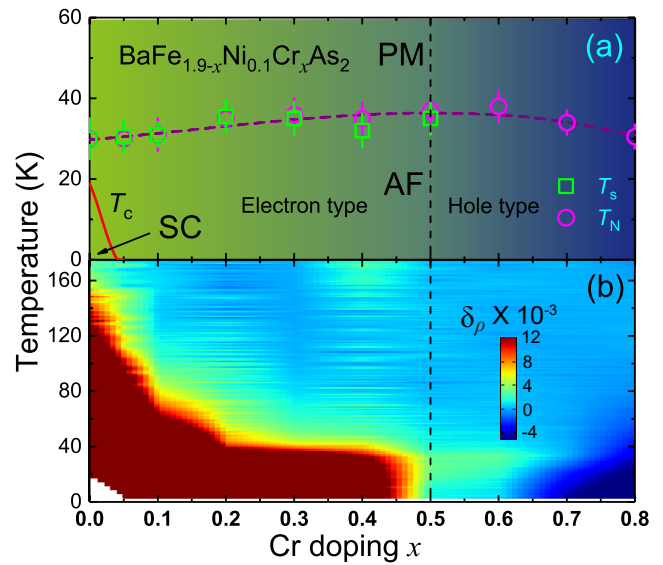}
\caption{(Color online) Phase diagram and in-plane resistivity anisotropy of BaFe$_{1.9-x}$Ni$_{0.1}$Cr$_{x}$As$_{2}$. (a) The PM, AF and SC mark the region of paramagnetic, antiferromagnetic, superconducting phases defined by $T_s$, $T_N$ and $T_c$, respectively. Here $T_s$ and $T_N$ were measured by neutron diffraction in our previous work on the naturally twinned samples \cite{dgong2018}. (b) The gradient color maps the in-plane resistivity anisotropy $\delta_{\rho}$ measured on detwinned samples. The vertical dashed line divides the regions for electron-type and hole-type charge carriers as determined by the sign of Hall and Seebeck coefficients \cite{dgong2018}.
 }
\end{figure}

\section{Experiment details}

\begin{center}
\begin{figure*}[t]
\includegraphics[width=1\textwidth]{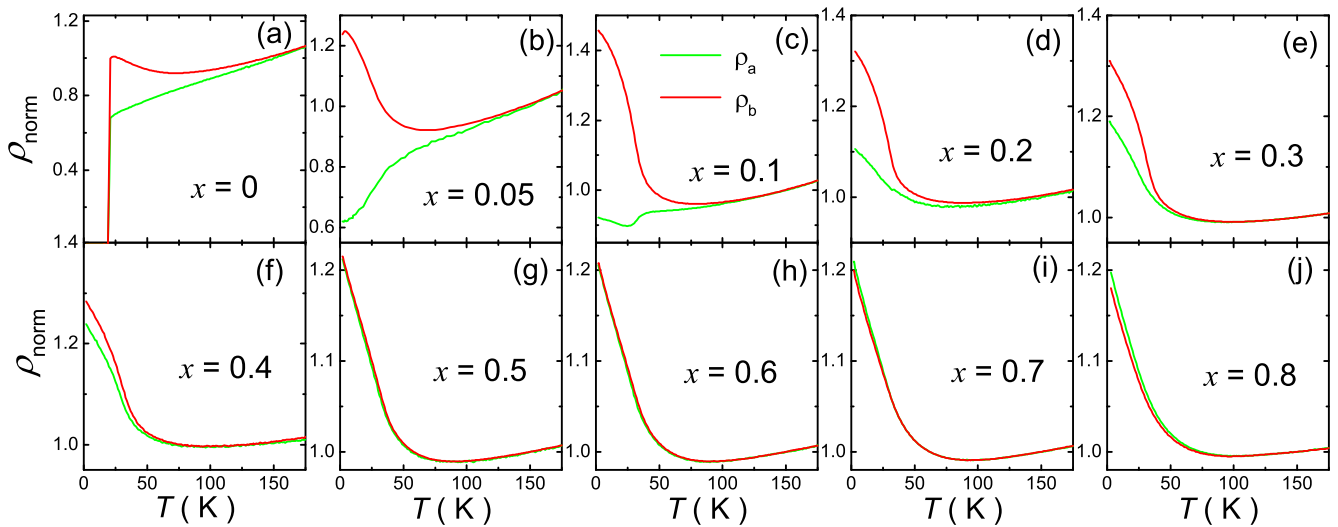}
\caption{(Color online) In-plane resistivity anisotropy of BaFe$_{1.9-x}$Ni$_{0.1}$Cr$_{x}$As$_{2}$ under uniaxial pressure. Here $\rho_b$ is the in-plane resistivity along the direction of uniaxial pressure, and $\rho_a$ is the in-plane resistivity perpendicular to the direction of uniaxial pressure. For easy comparison, each curve is normalized by its resistivity at 150 K, there is no resistivity anisotropy above this temperature.
 }
\end{figure*}
\end{center}

\begin{figure}[t]
\includegraphics[width=0.45\textwidth]{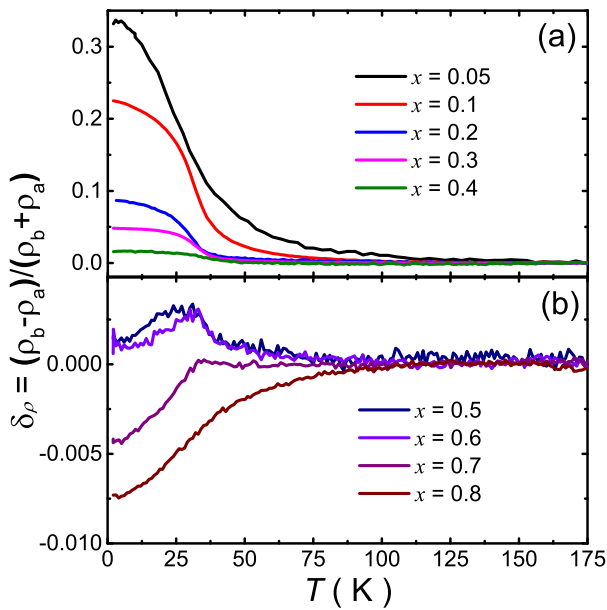}
\caption{(Color online) Temperature dependence of the in-plane resistivity anisotropy $\delta_{\rho}$ from $x=0.05$ to 0.8. (a) In electron-type compounds, $\delta_{\rho}$ gets weaker but keeps positive when increasing Cr doping. (b) In hole-type compounds, $\delta_{\rho}$ is very weak and becomes negative when $x\geq0.7$.
 }\label{fig3}
\end{figure}

High-quality single crystals of BaFe$_{1.9-x}$Ni$_{0.1}$Cr$_{x}$As$_{2}$ were grown by the self-flux method \cite{rzhang2014,rzhang2015,dgong2018,hqluo2008,ychen2011,txie2017,twang2020b}, the characterization results of our sample can be found in previous reports \cite{rzhang2014,dgong2018}. The crystalline directions of our sample were determined by a X-ray Laue camera (\textit{Photonic Sciences}) in backscattering mode with incident beam along $c-$axis. After that, the crystals were cut into rectangle shapes (typical sizes: 1 mm $\times$ 2 mm) by a wire saw under the directions [1, 0, 0] $\times$ [0, 1, 0] in orthorhombic lattice notation ($a=b= 5.6$ \AA). By applying a uniaxial pressure around 10 MPa, the crystal can be fully detwinned at low temperature, where the direction of pressure was defined as $b$ direction, and the pressure-free direction was defined as $a$ direction \cite{xlu2014,wzhang2016,ysong2015,xlu2018,xlu2016,dwtam2019,dwtam2020,pliu2020}. The in-plane resistivity ($\rho_{a,b}$) was measured by the standard four-probe method with the Physical Property Measurement System (PPMS) from Quantum Design. To compare the temperature dependence of resistivity at different directions, we normalized the resistivity $\rho_{a,b} (T)$ data at 150 K for each sample. The in-plane resistivity anisotropy was defined by $\delta_{\rho}=(\rho_b-\rho_a)/(\rho_b+\rho_a)$ {same as other literatures \cite{jhchu2010,matanatar2010,jjying2011}.

ARPES experiments were performed at beamline 10.0.1 of the Advanced Light Source and beamline 5-4 of the Stanford Synchrotron Radiation Light source with R4000 electron analyzers. The angular resolution was 0.3 degree, and the total energy resolution was 15 meV. All samples were cleaved \emph{in-situ} at 10 K and measured in ultra-high vacuum with a base pressure lower than 4 $\times 10^{-11}$ Torr. We note that we used twinned samples without uniaxial pressure for the ARPES experiments. INS experiments were carried out at two thermal triple-axis spectrometers: PUMA at Heinz Maier-Leibnitz Zentrum (MLZ) \cite{sobolev2015}, Germany, and TAIPAN at Australian Centre for Neutron Scattering (ACNS) \cite{danilkin2009}, ANSTO, Australia. The wave vector $\bf Q$ at ($q_x$, $q_y$, $q_z$) was defined as $(H,K,L) = (q_xa/2\pi, q_yb/2\pi, q_zc/2\pi)$ in reciprocal lattice units (r.l.u.) using the orthorhombic lattice parameters $a \approx b=5.6$ \AA, and $c \approx 13$ \AA. All measurements were done with a fixed final energy $E_f=14.8$ meV, and double focusing monochromator and analyzer using pyrolytic graphite crystals. To gain better signal-noise ratio, 8 pieces of rectangularly cut crystals (typical sizes: 7 mm $\times$ 8 mm $\times$ 0.5 mm) were assembled in a detwinned device made by aluminum and springy gaskets \cite{xlu2014,wzhang2016,ysong2015,xlu2018}. To reach both $Q=(1, 0, 1)$ and $Q=(0, 1, 1)$, the sample holder was designed to easily rotate by 90 degrees, thus the scattering plane can switch from $[H, 0, 0] \times [0, 0, L]$ to $[0, K, 0] \times [0, 0, L]$. The total mass of the crystals used in INS experiments was about 2 grams from each sample set of $x=0.05$ and $x=$0.5. Time-of-flight neutron scattering experiments were carried out on the same sample sets at 4SEASONS spectrometer (BL-01) at J-PARC \cite{nakamura2009,kajimoto2011}, Tokai, Japan, with multiple incident energies $E_i=250, 73, 34, 20$ meV, $k_i$ parallel to the $c$ axis, and chopper frequency $f=250$ Hz. The data were only corrected by the efficiency of detectors from the incoherent scattering of vanadium with white beam. As we were comparing two samples with similar mass under the same measured conditions at the same spectrometer, it was not necessary to do the vanadium normalization with mono beam. The data was analyzed by the Utsusemi and MSlice software packages \cite{inamura2013,mslice}.

\section{Results and discussions}

We first present the resistivity results in Fig. 1, Fig. 2 and Fig. 3. Apparently, the in-plane resistivity anisotropy show a strong dependence on the Cr doping level. In the Cr free sample BaFe$_{1.9}$Ni$_{0.1}$As$_{2}$, difference between $\rho_{a}$ and $\rho_{b}$ presents above the superconducting transition temperature $T_c=20$ K, where $\rho_{a}$ is metallic and $\rho_{b}$ is semiconducting-like with a upturn at low temperature (namely $\rho_{a}<\rho_{b}$) [Fig. 2(a)]. The superconductivity is completely suppressed at $x=0.05$, there is a dramatic difference between $\rho_{a}$ and $\rho_{b}$ with a anisotropy $\delta_{\rho}$ persisting to about $T=110$ K [Fig. 2(b)]. By further increasing Cr doping, both $\rho_{a}$ and $\rho_{b}$ become semiconducting-like even insulating-like above $x=0.1$, and the resistivity anisotropy gets weaker and weaker, until it nearly disappears at $x=0.5$ and 0.6 compounds [Fig. 2(c) - (h)]. For those high doping compounds $x=0.7$ and 0.8, it seems that $\delta_{\rho}$ changes sign with $\rho_{a}>\rho_{b}$ at low temperatures [Fig. 2(i) and (j)]. To clearly compare the resistivity anisotropy upon Cr doping, we plot $\delta_{\rho}$ as gradient color mapping in Fig. 1(b) and show its detailed temperature dependence in Fig. 3. Interestingly, the sign of $\delta_{\rho}$ is also related to the type of charge carriers. $\delta_{\rho}$ keeps strong and positive in the electron-type compounds but changes to negative and weak ($< 1\%$) in the hole-type compounds [Fig. 1(b) and Fig. 3(b)]. This is consistent with the results in the electron doped BaFe$_{2-x}$(Ni, Co)$_x$As$_2$ and the hole doped Ba$_{1-x}$K$_x$Fe$_2$As$_2$, Ca$_{1-x}$Na$_x$Fe$_2$As$_2$, BaFe$_{2-x}$Cr$_x$As$_2$ systems \cite{jhchu2010,matanatar2010,jjying2011,hman2015,blomberg2013,jqma2013,kobayashi2015,ishida2020}. However, in those cases, the onset temperature of $\delta_{\rho}$ decreases with the structural transition temperature $T_s$ when increasing the doping level from the non-superconducting parent compounds to optimally doped superconducting compounds. Here in BaFe$_{1.9-x}$Ni$_{0.1}$Cr$_x$As$_2$ system, both $T_N$ and $T_s$ are actually within the range $32 \sim 35$ K for all probed dopings \cite{dgong2018}, but the onset temperature of $\delta_{\rho}$ still extends to high temperatures, it is then strongly suppressed by Cr doping [Fig. 3(a)]. In those hole-type compounds, $\delta_{\rho}$ shows a peak feature (for $x=0.5$ and 0.6) or a kink (for $x=0.7$ and 0.8) responding to the magnetic and structural transitions [Fig. 3(b)].  The non-monotonic behavior of $\delta_{\rho}$ may come from the competition between the scattering from hole bands and electron bands, similar behaviors were observed in the nematic susceptiblity of the Cr doped BaFe$_{2}$(As$_{1-x}$P$_x$)$_{2}$ system \cite{wzhang2019}.

\begin{figure}[t]
\includegraphics[width=0.45\textwidth]{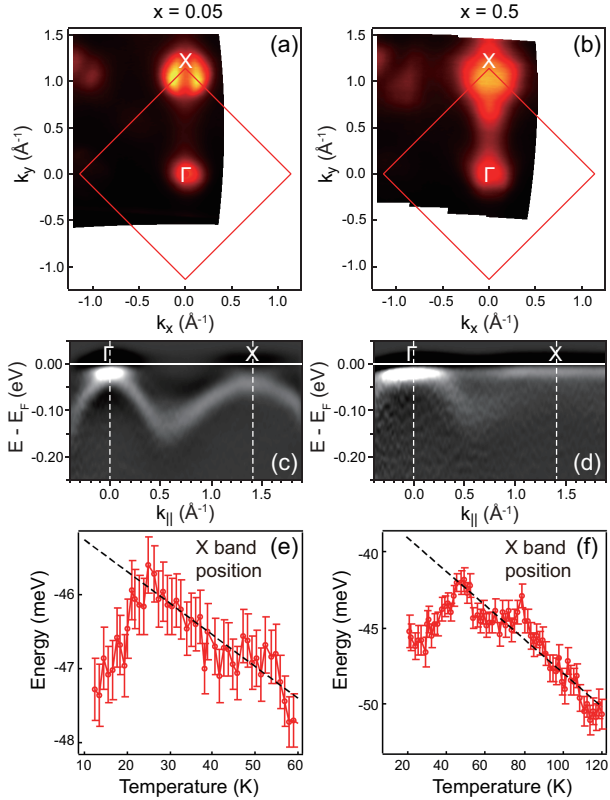}
\caption{(Color online) ARPES results on $x=0.05$ (left) and $x=0.5$ (right) compounds. (a)-(b) The measured Fermi surfaces around the $\Gamma$ and $X$ points. (c)-(d) Band dispersions along the high symmetry direction $\Gamma$-$X$ obtained from the second derivatives in the energy direction. (e)-(f) Temperature dependence of the fitted band position from the $X$ point. All dashed lines are guides for eyes.
}
\end{figure}

\begin{figure*}[t]
\includegraphics[width=0.95\textwidth]{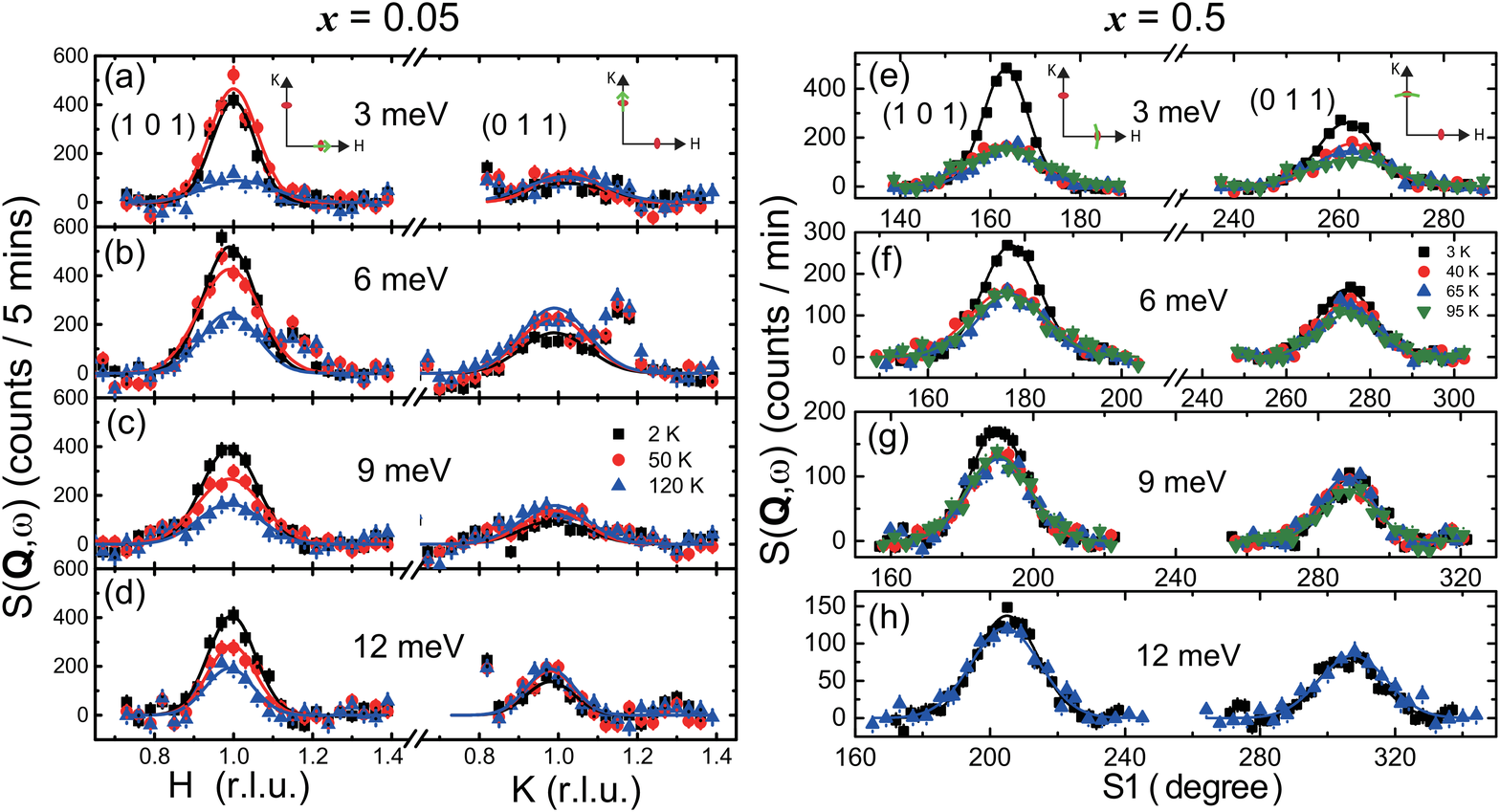}
\caption{(Color online) Inelastic neutron scattering results on the spin excitations of uniaxially detwinned samples for $x=0.05$ (left) and $x=0.5$ (right) compounds measured by two triple-axis spectrometers TAIPAN and PUMA. We compared the constant-energy scans ($Q-$scans along $[H, 0, 1]$ or $[0, K, 1]$, $S1$ rocking scans at $Q=(1, 0, 1)$ or (0, 1, 1)) for $E=$ 3, 6, 9, 12 meV, respectively. All data are corrected by a linearly $Q-$dependent background, and the solid lines are gaussian fittings. The spurious signals in 6 meV data are ignored.
}
\end{figure*}

\begin{figure*}[t]
\includegraphics[width=0.9\textwidth]{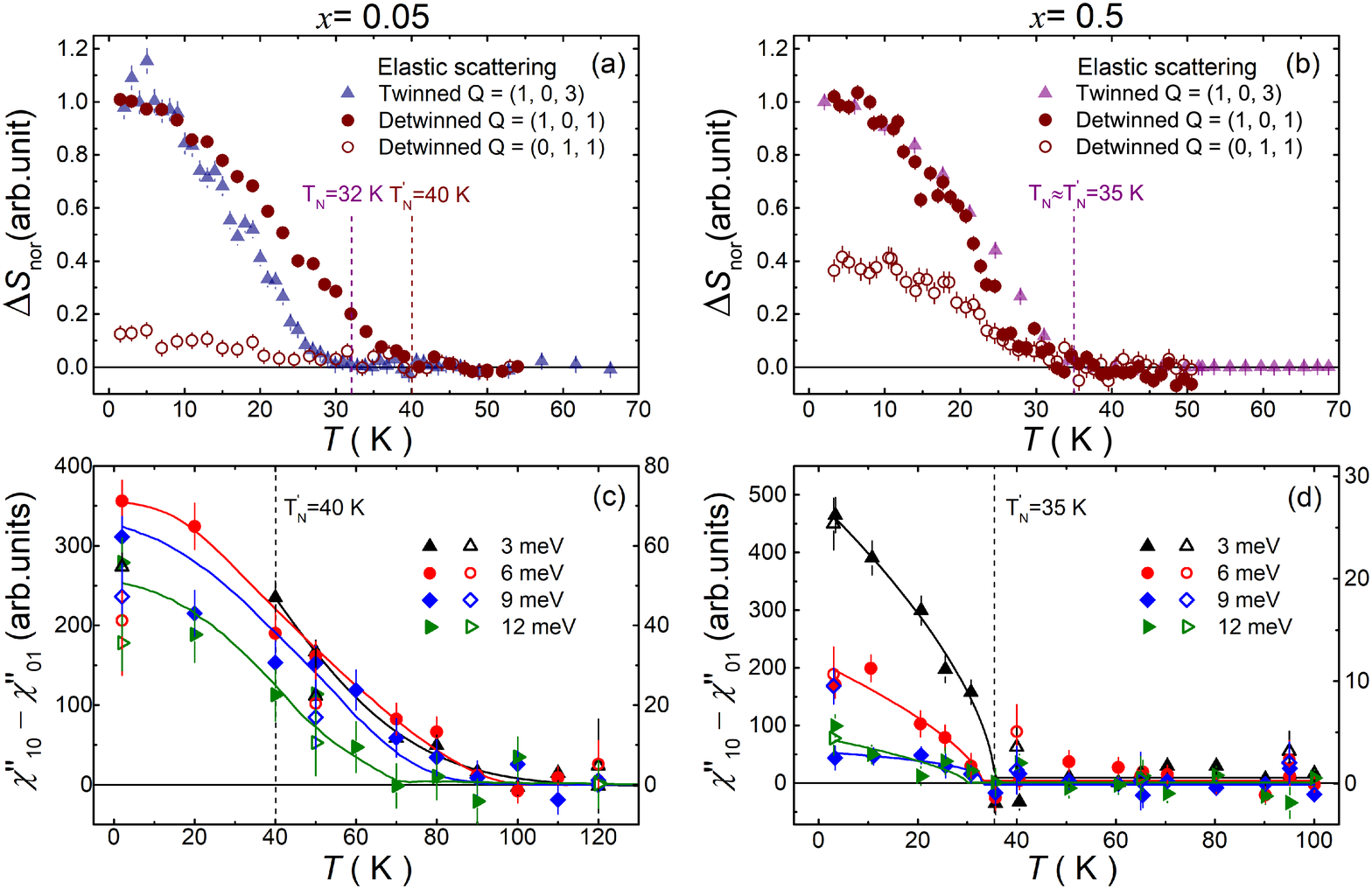}
\caption{(Color online) The order parameter of antiferromagnetism and spin nematicity $\chi^{\prime\prime}_{10}-\chi^{\prime\prime}_{01}$ for $x=0.05$ and $x=0.5$ compounds.(a) and (b) The magnetic order parameters measured at $Q=(1, 0, 3)$ on twinned samples, $Q=(1, 0, 1)$ and $Q=(0, 1, 1)$ on detwinned samples by elastic neutron scattering. All data are subtracted by the normal state background and normalized by the intensity at base temperature for $Q=(1, 0, 3)$  or $Q=(1, 0, 1)$. (c) and (d) Spin nematicity measured by inelastic neutron scattering. The solid symbols are the differences of local susceptibility $\chi^{\prime\prime}$ between $Q=(1, 0, 1)$ and $Q=(0, 1, 1)$ (left y-axis), and the open symbols are similar but obtained by integrating the constant-energy scans in Fig. 5 corrected by the Bose population factor (right y-axis). The vertical dash lines mark the magnetic transition temperature $T_N$ on twinned samples and $T^{\prime}_N$ on detwinned samples. All solid lines are guides to eyes.
 }\label{fig6}
\end{figure*}

Next, we focus on the electronic structure and the spin excitations in two typical dopings $x=0.05$ with $T_N=32$ K and $x=0.5$ with $T_N=35$ K. The Fermi surface topology and band structure measured by ARPES on naturally twinned samples are shown in Fig. 4. From the Fermi surface mapping in Fig. 4 (a) and (b), we can find typical hole pockets around the zone center $\Gamma$ point. Near the $X$ point, an electron pocket is observed for $x=0.05$. For $x=0.5$, however, the Fermi surface resembles that of the hole-doped (Ba,K)Fe$_2$As$_2$\cite{myi2014}. This is due to the hole doping introduced by the Cr substitution, which also introduces disorder directly in the Fe-planes, thus resulting in spectral features that appear broad \cite{myi2009}. Figures 4 (c) and (d) show the second energy derivatives of the spectral images along the high symmetry direction ($\Gamma$-$X$). Larger hole pockets can indeed be seen for $x=0.5$ compared to $x=0.05$. As has been demonstrated previously on BaFe$_2$As$_2$, NaFeAs, and FeSe, the onset of $T_s$ is associated with the onset of an observed anisotropic shift of the $d_{xz}$ and $d_{yz}$ orbital-dominated bands where the $d_{xz}$ band shifts down and the $d_{yz}$ band shifts up\cite{myi2011,myi2014,myi2012,myi2019,yzhang2012}. This shift is most prominently observed near the X point of the Brillouin zone. Moreover, such band splitting as measured on uniaxially-strained crystals can be observed above $T_s$ in the presence of this symmetry-breaking field. On a structurally twinned crystal, the anisotropic band shifts would appear in the form of a band splitting due to domain mixing. While we do not observe clearly the band splitting as shown in Fig. 4(c)-(d), we can clearly observe the lower branch with dominant intensity that shifts with temperature. This can be understood as the lower $d_{xz}$ band. We can fit the energy position of the band extracted from the X point and plot as a function of temperature. The temperature evolution clearly identifies a temperature scale associated with an onset of the band shift \cite{myi2014,myi2012,myi2019}. As shown in Fig. 4 (e) and (f), the $X$ band shifts at low temperature $T\approx 25$ K for $x=0.05$ and $T\approx 45$ K for $x=0.5$, respectively, closing to their structural or magnetic transition temperatures. We do note that while we cannot conclusively state that this represents the orbital anisotropy, the behavior we observe here on these twinned crystals is consistent with the expectation of the onset of orbital anisotropy\cite{myi2011,myi2017,ysong2015}. We note here that the observed onset temperature of band splitting is close to the $T_s$ (or $T_N$), in contrast to the much higher onset in the resistivity anisotropy shown in Fig. 3 measured on a strained crystal.

We then turn to search the connection between the resistivity anisotropy and the spin excitation anisotropy. The first evidence of spin nematicity was observed in BaFe$_{2-x}$Ni$_{x}$As$_{2}$ ($x=0$, 0.065, 0.085, 0.10, 0.12) \cite{xlu2014,wzhang2016,ysong2015,xlu2018}, where BaFe$_{1.9}$Ni$_{0.1}$As$_{2}$ is the starting compound of this study. Low-energy spin excitations are measured on the detwinned BaFe$_{1.9-x}$Ni$_{0.1}$Cr$_x$As$_2$ ($x=0.05$ and 0.5) samples by INS experiments using two triple-axis spectrometers. The results of constant-energy scans at $E=$ 3, 6, 9, and 12 meV are summarized in Fig. 5.  With convenient design of the detwinned device and sample holder, we can easily perform constant-energy scans ($Q-$scans) either along $[H, 0, 1]$ or $[0, K, 1]$ direction after rotating the whole sample set by 90 degree. For $x=0.5$ sample, we instead do the $S1$ rocking scans at $Q=(1, 0, 1)$ and (0, 1, 1). It should be noticed that the N\'{e}el temperature $T_N$ is slightly enhanced by the applied uniaxial pressure in $x=0.05$ sample from 32 K to 40 K (so does $T_s$), but does not change for $x=0.5$ sample ($T_N \approx T^{\prime}_N=35$ K) [Fig. 6(a) and (b)]. Such effect has been detected in the BaFe$_{2-x}$(Ni, Co)$_{x}$As$_{2}$ system \cite{dwtam2017}. The detwinned ratio can be estimated by comparing the integrated intensities of magnetic Bragg peak between $Q=(1, 0, 1)$ and $Q=(0, 1, 1)$ positions, which is about 10 : 1 for the $x=0.05$ samples, and 4 : 1 for the $x=0.5$ samples, respectively. Such large ratio means successful detwin for both sample sets. At the first glance, it is very clear for the difference of the spin excitations between $Q=(1, 0, 1)$ and $Q=(0, 1, 1)$ especially at low temperatures, which could be attributed to the spin Ising-nematic correlations (so called spin nematicity). After warming up to high temperatures, the spin excitations at $Q=(1, 0, 1)$ decrease and become nearly identical to those at $Q=(0, 1, 1)$. The nematic order parameter for the spin system can be approximately represented by $\chi^{\prime\prime}_{nematic}=\chi^{\prime\prime}_{10}-\chi^{\prime\prime}_{01}$, in which $\chi^{\prime\prime}_{10}$ (or $\chi^{\prime\prime}_{01}$) is the local spin susceptibility at $Q=(1, 0, 1)$ (or $Q=(0, 1, 1)$). Figure 6 (c) and (d) show the temperature dependence of $\chi^{\prime\prime}_{10}-\chi^{\prime\prime}_{01}$ for both compounds, where the Bose population factor is already corrected. We also plot the data (open symbols) obtained from the integrated intensity of those $Q-$scans in Fig. 5. For $x=0.05$ compound, the spin nematicity decreases slightly upon increasing energy and terminates well above $T^{\prime}_N= 40$ K [Fig. 6(c)] \cite{dgong2018}. For the lowest energy we measured (3 meV), the onset temperature of spin nematicity is about 110 K, similar to the in-plane resistivity anisotropy in Fig. 3(a). The results for $x=0.5$ compound show markedly differences, where $\chi^{\prime\prime}_{10}-\chi^{\prime\prime}_{01}$ quickly decreases both with energy and temperature, and the onset temperature is around $T^{\prime}_N= 35$ K [Fig. 6(d)]. No spin anisotropy can be detected above 40 K both for $Q-$scans and energy scans, this is also consistent with the very weak in-plane resistivity anisotropy for $x=0.5$ [Fig. 3(b)]. The spin nematic theory predicts that the nematic fluctuations enhance both the intensity and the correlation length of spin excitations at ($\pi$, 0) but suppress those at (0, $\pi$) even above $T_s$. This was firstly testified in the detwinned BaFe$_{1.935}$Ni$_{0.065}$As$_2$ and can be also seen here in Fig. 5 \cite{wzhang2016}. Although the peak intensities at $Q=(1, 0, 1)$ seem stronger than those at $Q=(0, 1, 1)$ in Fig. 5(g) and (h), the peak width is smaller, and the integrated intensity of the $Q$-scans are closed to each other. The above results of spin nemacticity in BaFe$_{1.9-x}$Ni$_{0.1}$Cr$_x$As$_2$ ($x=0.05$ and 0.5) resemble to those in BaFe$_{2-x}$Ni$_{x}$As$_{2}$, where spin excitations at low energies change from $C_4$ to $C_2$ symmetry in the tetragonal phase at temperatures approximately corresponding to the onset of the in-plane resistivity anisotropy.

\begin{figure*}[t]

\includegraphics[width=0.9\textwidth]{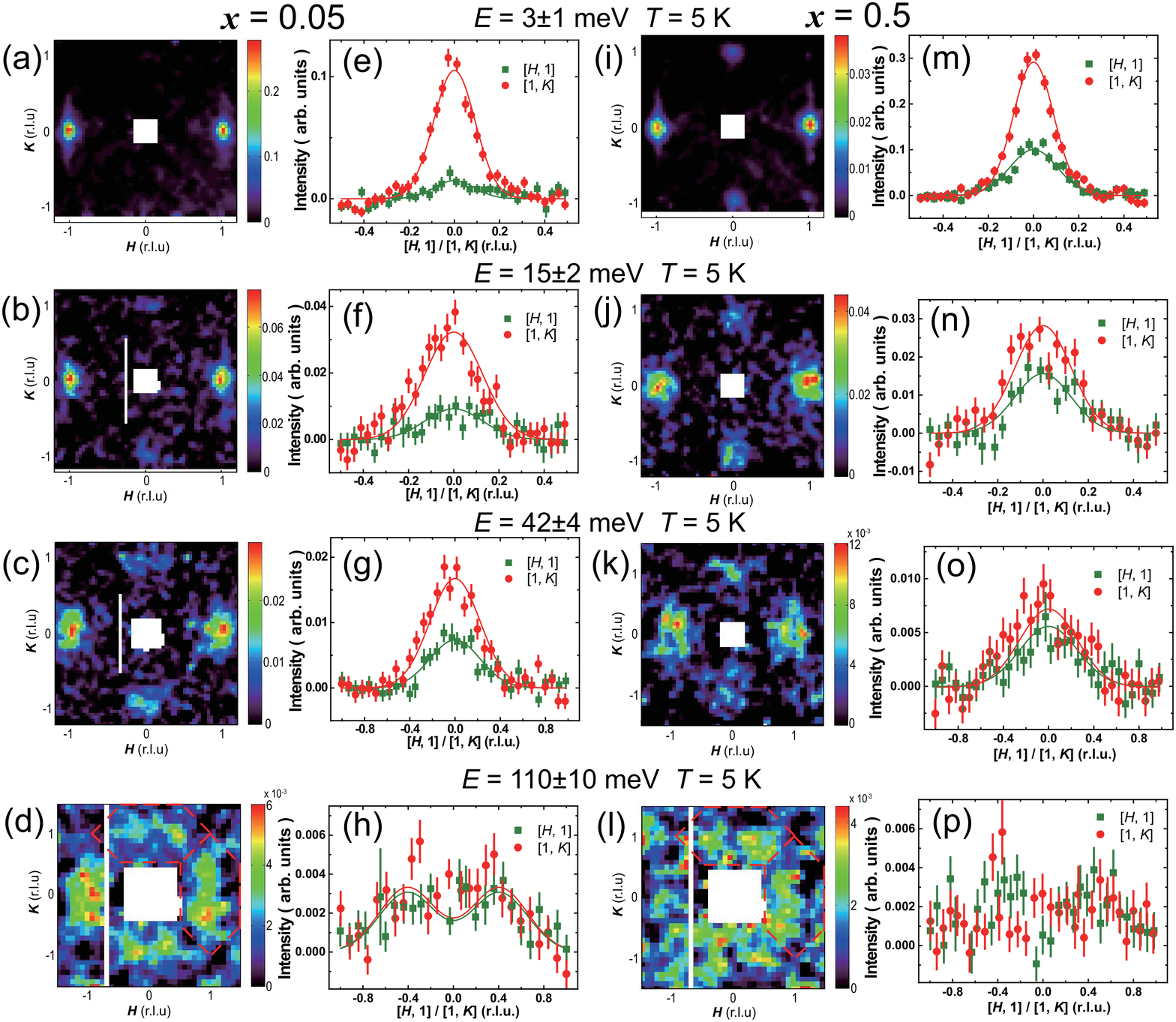}
\caption{(Color online) Inelastic neutron scattering results on the spin excitations of the uniaxially detwinned samples for $x=0.05$ (left) and $x=0.5$ (right) compounds measured at $T=5$ K by time-of-flight spectrometer 4SEASONS. All data are presented both in 2D slices for $[H, K]$ plane and 1D cuts along $[H, 0]$ or $[0, K]$ at typical energy windows $E=3\pm1,15\pm2,42\pm4,110\pm10$ meV. The solid lines are gaussian fittings guiding for eyes, which are not shown in panel (p) due to poor data quality. The dashed diamonds in panel (d) and (l) illustrate the integrated Brillouin zone for spin excitations around $Q=(1, 0)$ and $Q=(0, 1)$, respectively.
}
\end{figure*}

\begin{figure*}[t]
\includegraphics[width=0.9\textwidth]{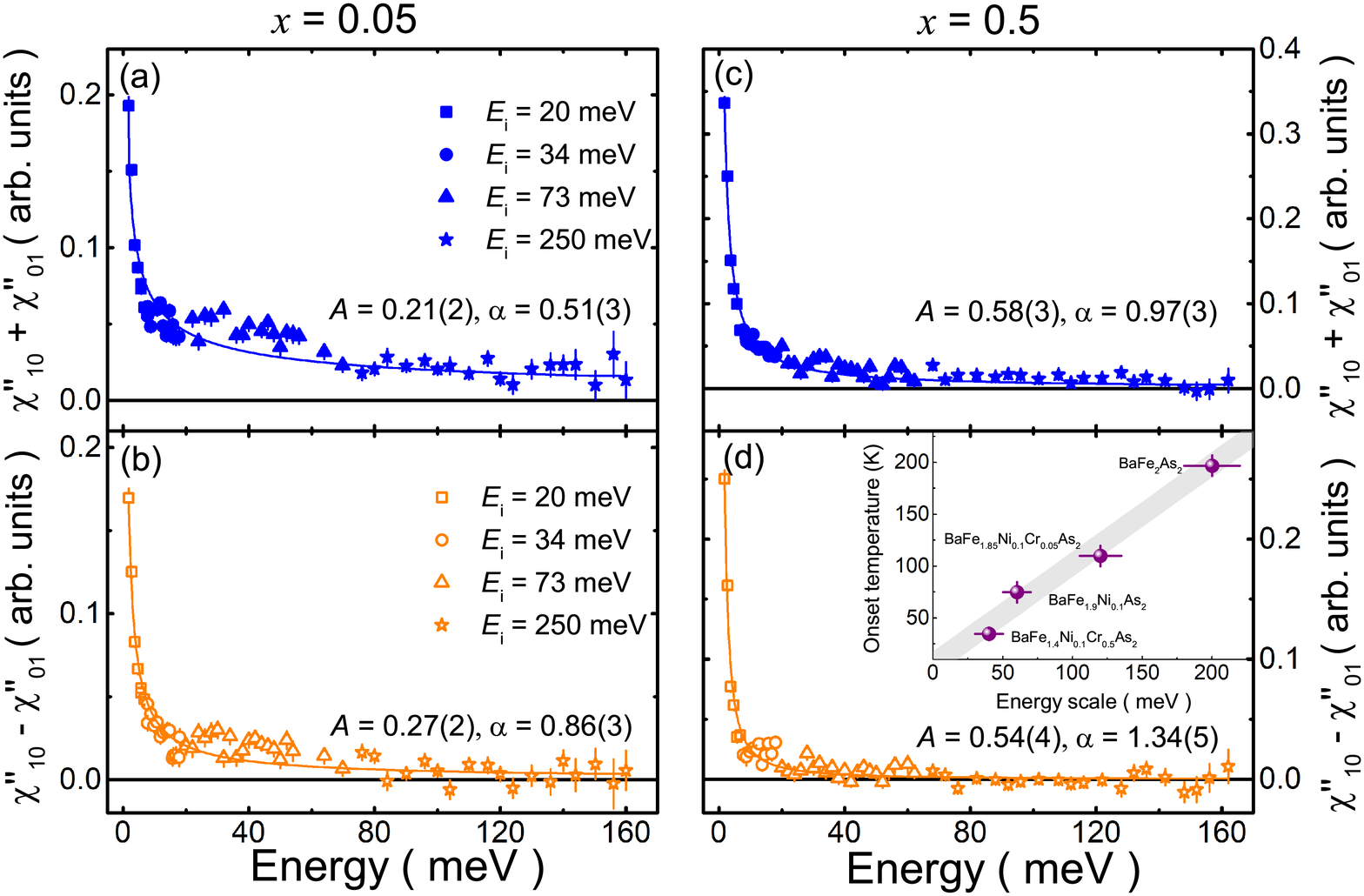}
\caption{(Color online) Energy dependence of the total spin fluctuations $\chi^{\prime\prime}_{10}+\chi^{\prime\prime}_{01}$ and the spin nematicity $\chi^{\prime\prime}_{10}-\chi^{\prime\prime}_{01}$ of uniaxially detwinned samples for $x=0.05$ (left) and $x=0.5$ (right) compounds. Different symbols correspond to different incident energy in the measurements. Both of $\chi^{\prime\prime}_{10}+\chi^{\prime\prime}_{01}$  and $\chi^{\prime\prime}_{10}-\chi^{\prime\prime}_{01}$ can be fitted with a power-law dependence on the energy, $\sim A/E^{\alpha}$, where the amplitude $A$ and exponent $\alpha$ are listed in each panel. Inset of panel (d) shows the correlation between the highest energy and the onset temperature at low energy of $\chi^{\prime\prime}_{10}-\chi^{\prime\prime}_{01}$.
}
\end{figure*}

Moreover, INS experiments on detwinned BaFe$_{2}$As$_{2}$ and BaFe$_{1.9}$Ni$_{0.1}$As$_{2}$ suggest the spin anisotropy can persist to very high energy \cite{ysong2015,xlu2018}, even in the later case the splitting of the $d_{xz}$ and $d_{yz}$ bands nearly vanishes \cite{myi2017}. To quantitatively determine the energy dependence of spin excitation anisotropy, we have performed time-of-flight INS experiments on the uniaxially detwinned BaFe$_{1.9-x}$Ni$_{0.1}$Cr$_x$As$_2$ ($x=0.05$ and 0.5), the results are shown in Fig. 7 and Fig. 8. It should be noted that for such experiments, the energy transfer is always in coupled with $L$ due to $k_i \parallel c$ \cite{nakamura2009,kajimoto2011}. The two-dimensional (2D) energy slices and one-dimensional (1D) cuts along $[H, 0]$ and $[0, K]$ at various energies are presented in Fig. 7. Indeed the spin excitations are twofold symmetric below 100 meV for both compounds. The spin excitations at $E=3$ meV, $Q=(0, \pm 1)$ are very weak in $x=0.05$ compound, then continuously increase upon energy and become nearly the same as $Q=(\pm 1, 0)$ around 110 meV. For $x=0.5$ compound, although the spin excitations at $Q=(0, \pm 1)$ can be initially observed at $E=3$ meV, the spin anisotropy still exists at 15 meV, then disappears above 42 meV. To further compare the spin excitations in both compounds, we have calculated the total spin fluctuations $\chi^{\prime\prime}_{10}+\chi^{\prime\prime}_{01}$ and the spin nematicity $\chi^{\prime\prime}_{10}-\chi^{\prime\prime}_{01}$ from the integrated intensity marked by the dashed diamonds in Fig. 7 (d) and (l). In principle, the local dynamic susceptibility $\chi^{\prime\prime}$ can be estimated from the integration outcome of the spin excitations within one Brillouin zone, here $\chi^{\prime\prime}$ can simply be calculated through dividing the integration signal in the $Q=$ (0, 0), (1, 1), (2, 0), (1, -1) boxes, giving the diamond shape integration zone \cite{pdai2015}. The total spin susceptibility $\chi^{\prime\prime}_{10}+\chi^{\prime\prime}_{01}$ in $x=0.5$ compound is stronger than that in $x=0.05$, but decays much quickly with energy [Fig. 8 (a) and (c)]. The spin nematicity $\chi^{\prime\prime}_{10}-\chi^{\prime\prime}_{01}$ apparently has different energy scale for two compounds, where it is about 120 meV for $x=0.05$ but only 40 meV for $x=0.5$, respectively. The energy scale of $\chi^{\prime\prime}_{10}-\chi^{\prime\prime}_{01}$ in the superconducting compound BaFe$_{1.9}$Ni$_{0.1}$As$_{2}$ is 60 meV \cite{ysong2015}, and for the parent compound BaFe$_2$As$_2$ is about 200 meV up to the band top of the spin waves \cite{xlu2018}. These facts lead to a possible linear correlation between the highest energy and the onset temperature of spin nematicity at low energy [inset of Fig. 8(d)]. Within the measured energy range, both $\chi^{\prime\prime}_{10}+\chi^{\prime\prime}_{01}$ and $\chi^{\prime\prime}_{10}-\chi^{\prime\prime}_{01}$ can be fit with a power-law dependence on the energy, $\sim A/E^{\alpha}$, where the amplitude $A$ and exponent $\alpha$ are listed in each panel of Fig. 8. Indeed, the larger value of $\alpha$ for $x=0.5$ in comparison to that for $x=0.05$ suggests faster decay with energy both for the spin fluctuations and the spin nematicity. Similar fitting on the results of BaFe$_{1.9}$Ni$_{0.1}$As$_{2}$ gives parameters in between them \cite{ysong2015}. Although the low energy data below 10 meV may be affected by the $L$-modulation of spin excitations, and by the superconductivity in BaFe$_{1.9}$Ni$_{0.1}$As$_{2}$, the similar quantum critical behavior both for $\chi^{\prime\prime}_{10}+\chi^{\prime\prime}_{01}$ and $\chi^{\prime\prime}_{10}-\chi^{\prime\prime}_{01}$ in these three compounds is expected by the Ising-nematic scenario \cite{ysong2015,xlu2014,xlu2018,wzhang2016}.

In our previous neutron diffraction results on the BaFe$_{1.9-x}$Ni$_{0.1}$Cr$_{x}$As$_{2}$ system, the Cr dopings have limited effects on the magnetically ordered temperature $T_N$ but significantly enhance the effective ordered moment $m$ by reaching a maximum value at $x=0.5$ \cite{dgong2018}. The N\'{e}el temperature $T_N$ is mostly determined by the local magnetic coupling related to the local FeAs$_4$ tetrahedron structure. The evolution of ordered moment probably induced by the changes of the density of states and the orbital angular momentum from itinerant electrons on the Fermi surfaces. The Cr doping introduces both local distortion on the lattices and hole doping on the Fermi pockets, yielding a non-monotonic change of the conductivity of charge carriers. As shown in Fig. 2, the low-temperature upturn of resistivity is enhanced by Cr doping first but then weakens in those hole-type compounds. Among these dopings, $x=0.5$ has the most insulating-like behavior, and thus strongly localized charge carriers and maximum ordered moment, but its spin nematicity quickly drops down both for the temperature and energy dependence. In contrast to the magnetically ordered strength, both the structural transition temperature $T_s$ and the lattice orthorhombicity $\delta=(a-b)/(a+b)$ are nearly Cr doping independent \cite{dgong2018}. This means the static nematic order is also nearly Cr independent in this system, as opposed to the case for dynamic nematic fluctuations.

The nature of iron-based superconductor can be theoretically described as a magnetic Hund's metal, in which the strong interplay between the local spins on Fe atoms and the itinerant electrons on Fermi surfaces gives correlated electronic states \cite{zpyin2011,georges2013}. Indeed, time-of-flight INS experiments on the detwinned BaFe$_{2}$As$_{2}$ suggest the spin waves in parent compound are preferably described by a multi-orbital Hubbard-Hund model based on the itinerant picture with moderate electronic correlation effects, instead of a Heisenberg model with effective exchange couplings from local spins.  Upon warming up to high temperatures, the intensities of spin excitation anisotropy decrease gradually with increasing energy and finally cut off at a energy away from the band top of spin waves \cite{xlu2018}. Therefore, the energy scale of spin nematicity sets an upper limit for the characteristic temperature for the nematic spin correlations, as well as the onset temperature of resistivity anisotropy. Here by adding up the results on the in-plane anisotropies of resistivity, orbital energy and spin excitations in BaFe$_{1.9-x}$Ni$_{0.1}$Cr$_{x}$As$_{2}$, they clearly suggest that the electronic nematicity is intimately related to the spin dynamics, which seems consistent with the Hund's metal picture. Specifically, by doping Cr to suppress the superconductivity in BaFe$_{1.9}$Ni$_{0.1}$As$_{2}$ makes the charge carriers initially localized with enhanced electron correlations \cite{dgong2018}, which may enhance the electronic correlations by increasing the intra- and inter-orbital onsite repulsion $U$ as well as the Hund's coupling $J_H$ \cite{xlu2018}, and thus gives rise to stronger spin excitations and larger spin anisotropy in Cr doping $x=0.05$ compound. Another effect is the lifting up of $d_{yz}$ and $d_{xy}$ along $\Gamma$-X direction to the Fermi level, which primarily contributes to the effective moments \cite{zpyin2011}. The orbital-weight redistribution triggered by the spin order suggests that the orbital degree of freedom are coupled to the spin degree of freedom \cite{mdaghofer2010}. By further increasing Cr doping to $x=0.5$, the localization effect is so strong that electron system becomes insulating at low temperature. In this case, the itinerant picture based on Hund's metal may not be applicable anymore. The low density of itinerant electrons weakens the nematic fluctuations and probably limits them inside the magnetically ordered state. In either case for $x=0.05$ or $x=0.5$, the band splitting does not directly correspond to the spin nematic correlations but only present below the nematic ordered temperature. This may attribute to the weak spin-orbit coupling in this system, as the spin anisotropy in spin space can only present at very low energies \cite{hluo2013}. In addition, our results can rule out the picture of local impurity scattering driven nematicity, since the impurity scattering from Cr substitutions is certainly stronger in the $x=0.5$ compound but it does not promote the nematic fluctuations.

\section{Conclusion}
In conclusion, we have extensively studied the in-plane resistivity anisotropy, orbital ordering and spin nematicity in a non-superconducting BaFe$_{1.9-x}$Ni$_{0.1}$Cr$_{x}$As$_{2}$ system. We have found the Cr doping strongly affect on the anisotropy of resistivity and spin excitations along with the itinerancy of charge carriers. While the onset temperatures of resistivity anisotropy and spin nematicity are similar and correlated to the energy scale of spin anisotropy, the orbital anisotropy shows a onset temperature irrelevant to them. These results suggest that the electronic correlations from the interplay between local moments and itinerant electrons is crucial to understand the nematic fluctuations, thus inspire the quest for the driven force of the electronic nematic phase in iron-pnictide superconductors.

{\bf Additional Requirements}
For additional requirements for specific article types and further information please directly refer to Huiqian Luo (hqluo@iphy.ac.cn).

{\bf Conflict of Interest Statement}
The authors declare that the research was conducted in the absence of any commercial or financial relationships that could be construed as a potential conflict of interest.

{\bf Author Contributions}
H. L. and D. G. proposed and designed the research. D. G., T. X., W. Z. and R. Z. contributed in sample growth and resistivity measurements. M. Y., M. W., S. M., M. H., D. L. and R. J. B. contributed to the ARPES measurements. D. G. and H. L. carried out the neutron scattering experiments with S. D., G. D., X. L., J. T. P., K. I., and K. K.. D. G., H. L., S. L. and P. D. analyzed the data. H. L., D. G. and M. Y. wrote the paper. All authors participated in discussion and comment on the paper.

{\bf Funding}
This work is supported by the National Key Research and Development Program of China (Grants No. 2018YFA0704200, No. 2017YFA0303100 and No. 2017YFA0302900), the National Natural Science Foundation of China (Grants No. 11822411, No. 11961160699, and No. 12061130200), the Strategic Priority Research Program (B) of the CAS (Grants No. XDB25000000 and No. XDB07020300) and K. C. Wong Education Foundation (GJTD-2020-01). H. Luo is grateful for the support from the Youth Innovation Promotion Association of CAS (Grant No. Y202001) and Beijing Natural Science Foundation (Grants No. JQ19002). M. Wang is supported by the National Natural Science Foundation of China (Grants No. 11904414, No.12174454), the Guangdong Basic and Applied Basic Research Foundation (No. 2021B1515120015), National Key Research and Development Program of China (No. 2019YFA0705702). Work at University of California, Berkeley and Lawrence Berkeley National Laboratory was funded by the U.S. Department of Energy (DOE), Office of Science, Office of Basic Energy Sciences, Materials Sciences and Engineering Division under Contract No. DE-AC02-05-CH11231 within the Quantum Materials Program (KC2202) and the Office of Basic Energy Sciences. The ARPES work at Rice University was supported by the Robert A. Welch Foundation Grant No. C-2024 (M. Y.).

{\bf Acknowledgements}
The authors thank the helpful discussion with Xingye Lu at Beijing Normal University and Yu Song at Zhejiang University. The neutron scattering experiments in this work are performed at thermal triple-axis spectrometer PUMA at Heinz Maier-Leibnitz Zentrum (MLZ), Germany, thermal triple-axis spectrometer TAIPAN at Australian Centre for Neutron Scattering (ACNS), Australian Nuclear Science and
Technology Organisation (ANSTO), Australia (Proposal No. P4263), and time-of-flight Fermi-chopper spectrometer 4SEASONS (BL-01) at the Materials and Life Science Experimental Facility of J-PARC (Proposal Nos. 2015A0005, 2016A0169). ARPES measurements were performed at the Advanced Light Source and the Stanford Radiation Lightsource, which are both operated by the Office of Basic Energy Sciences, U.S. DOE.

\end{document}